\documentclass[12pt]{article}

\usepackage{graphicx}

\def\beq   {\begin{equation}}
\def\eeq   {\end{equation}}
\def\beqd  {\begin{displaymath}}
\def\eeqd  {\end{displaymath}}
\def\bea {\begin{eqnarray}}
\def\eea {\end{eqnarray}}

\def\dr{\ifmmode{\overline{\rm DR}} \else{$\overline{\rm DR}$} \fi}
\def\ms{\ifmmode{\overline{\rm MS}} \else{$\overline{\rm MS}$} \fi}

\begin{document}


\vspace*{-1cm} 
\begin{flushright}
  TU-641 \\
  hep-ph/0112251
\end{flushright}

\vspace*{1.4cm}

\begin{center}

{\Large {\bf
Two-loop renormalization of $\tan\beta$ and its gauge dependence
} 
}\\

\vspace{10mm}

{\large Youichi Yamada}

\vspace{6mm}
\begin{tabular}{l}
{\it Department of Physics, Tohoku University, Sendai 980-8578, Japan}
\end{tabular}

\end{center}

\vspace{3cm}

\begin{abstract}
\baselineskip=15pt
Renormalization of two-loop divergent corrections to the vacuum 
expectation values ($v_1$, $v_2$) of the two Higgs doublets in the 
minimal supersymmetric standard model, 
and their ratio $\tan\beta=v_2/v_1$, is discussed 
for general $R_{\xi}$ gauge fixings. 
When the renormalized ($v_1$, $v_2$) are defined to give 
the minimum of the loop-corrected effective potential, it is shown that, 
beyond the one-loop level, the dimensionful parameters in the $R_{\xi}$ 
gauge fixing term generate gauge dependence of the renormalized $\tan\beta$. 
Additional shifts of the Higgs fields are necessary to realize 
the gauge-independent renormalization of $\tan\beta$. 
\end{abstract}

\vspace{20mm}

\newpage
\pagestyle{plain}
\baselineskip=15pt

Several extensions of the standard model have more than one 
Higgs boson doublets. 
For example, the minimal supersymmetric (SUSY) standard model 
(MSSM) \cite{mssm,gh} has two Higgs doublets
\beq
H_1=(H_1^0, H_1^-),\;\;\; H_2=(H_2^+, H_2^0).  \label{eq1} 
\eeq
Both $H_1^0$ and $H_2^0$ acquire the vacuum expectation 
values (VEVs) $v_i$ ($i=1,2$) which spontaneously break 
the SU(2) $\times$ U(1) gauge symmetry. 
$H_i^0$ are then expanded about the minimum of the Higgs potential as 
\beq
H_i^0 = \frac{v_i}{\sqrt{2}}+\phi_i^0. \label{eq2}
\eeq
$\phi_i^0$ are shifted Higgs fields with vanishing VEVs. 
I assume that CP violation in the Higgs sector is negligible 
and take $v_i$ as real and positive. 

A lot of physical quantities of the theory depend on the Higgs VEVs. 
In calculating radiative corrections to these quantities, 
the VEVs have to be renormalized. In the minimal standard model with 
only one Higgs doublet, the renormalization of the Higgs VEV is 
usually substituted by that of the weak boson 
masses \cite{sirlin,onshell}. However, this is not enough 
for extended theories with two or more Higgs VEVs. 
For example, the renormalization of $v_i$ 
in the MSSM is usually performed \cite{tanbrun,cpr,dabel} by specifying 
the weak boson masses, which are proportional to $v_1^2+v_2^2$, 
and the ratio $\tan\beta\equiv v_2/v_1$. Since $\tan\beta$ itself 
is not a physical observable, however, a lot of renormalization 
schemes for $\tan\beta$ have been proposed in the studies of the 
radiative corrections in the MSSM. Some of them are 
listed in Ref.~\cite{dpf}. 
In this letter, I concentrate on process-independent 
definitions of $\tan\beta$, which are given by the ratio of the 
renormalized VEVs $v_i$. I discuss the renormalization of the 
ultraviolet (UV) divergent corrections to $v_i$ and $\tan\beta$, 
working in the modified minimal subtraction schemes with dimensional 
reduction \cite{DR} (\dr scheme). The results are presented as 
the renormalization group equations (RGEs) for $v_i$ and $\tan\beta$. 
Since they are not physical observables, they may 
depend on the gauge fixing in general. I therefore 
investigate their gauge dependence in the general $R_{\xi}$ gauge 
fixing \cite{rxi}. 
Although I show the results for the MSSM, the results for the 
gauge dependence can be generalized for other models with 
two or more Higgs doublets. 

Even within the \dr scheme, there still remains an ambiguity of the way 
how to cancel the radiative shifts of the Higgs VEVs, $\Delta v_i$, 
by the one-point 
functions of $\phi^0_i$ by tadpole diagrams. One 
way is to cancel $\Delta v_i$ entirely 
by the shift of $\phi_i^0$. As a result, the tadpole contributions 
have to be added to all quantities which depend on $v_i$. 
The renormalized $v_i$ give the minimum of 
the tree-level Higgs potential and are just tree-level functions of 
the gauge-symmetric quadratic and quartic 
couplings in the Higgs potential. 
These $v_i$ are therefore independent of the gauge fixing 
parameters \cite{RGExi}. This renormalization scheme for $v_i$ is sometimes 
used \cite{degrassi,nielsenSM,mnuR,neumass} to show manifest gauge 
independence of physical quantities. However, since the running of 
$v_i$ in this scheme is very rapid \cite{RGEvsusy}, and the tadpole 
contributions appear in almost any corrections, this scheme is often 
inconvenient in practical calculations. 

Another, more popular way \cite{RGEvsusy,tanbrun,cpr,dabel,neumass} is to 
absorb $\Delta v_i$ by the shift of quadratic terms in the Higgs 
potential. The renormalized $v_i$ then give the minimum of 
the loop-corrected effective potential $V_{\rm eff}(H_1, H_2)$. 
This scheme is very convenient in practical calculation, 
because the explicit forms of the tadpole diagrams are necessary only 
for two-point functions of the Higgs bosons. 
However, the effective potential is generally dependent on the 
gauge fixing parameters \cite{veff1,veff2,veff3,cima}. 
The gauge dependence of the renormalized $v_i$ and 
their ratio $\tan\beta$ then might be a serious problem in 
calculating radiative corrections. 
I will therefore discuss the gauge dependence of the 
running $\tan\beta$ in this definition, in general $R_{\xi}$ gauges and 
to the two-loop order. 

The RGE for $v_i$ can be obtained from the UV divergent corrections to 
$v_i$-dependent masses or couplings of particles. 
For simplicity, I use the corrections to two quark masses $m_b$ 
and $m_t$, ignoring the masses of all other quarks and leptons. 
These mass terms are generated from the $b\bar{b}H_1$ and $t\bar{t}H_2$ 
Yukawa couplings, respectively, as 
\beq
L_{\rm int} = -h_b \bar{b}_R b_L (v_1/\sqrt{2}+\phi_1^0)
-h_t \bar{t}_R t_L (v_2/\sqrt{2}+\phi_2^0) 
+ {\rm h.c.} \label{eq3}
\eeq
The $R_{\xi}$ gauge fixing term takes the form 
\bea
L_{GF} &=& 
-\frac{1}{2\xi_Z} (\partial^{\mu}Z_{\mu} - \rho_Z G_Z)^2 
-\frac{1}{\xi_W} |\partial^{\mu}W^+_{\mu} -i\rho_W G_W^+|^2 \nonumber \\
&& -\frac{1}{2\xi_{\gamma}} (\partial^{\mu}\gamma_{\mu})^2
-\frac{1}{2\xi_g}\sum_{a=1}^8 (\partial^{\mu}g^a_{\mu})^2. \label{eq4}
\eea
The would-be Nambu-Goldstone bosons $G_V$ for $V=(Z,W)$ appear in 
Eq.~(\ref{eq4}). 
The parameters $\rho_V\equiv \xi_Vm_V$, where $m_V^2=g_V^2(v_1^2+v_2^2)/4$ 
($g_W^2=g_2^2$, $g_Z^2=g_2^2+g_Y^2$) are masses of $Z$ and $W^{\pm}$, 
are introduced in Eq.~(\ref{eq4}). 
This is to emphasize that the gauge symmetry 
breaking terms $\xi_Vm_V$ in $L_{GF}$, and also 
in the accompanied Fadeev-Popov ghost term, 
has very different nature from $v_i$ generated by the shifts (\ref{eq2}), 
as shown later. 
The terms $\rho_VG_V$ in Eq.~(\ref{eq4}) are expressed 
in the gauge basis (\ref{eq1}) of the Higgs bosons as 
\beq
\rho_Z G_Z = \xi_Z m_Z G_Z \equiv -{\sqrt{2}}{\rm Im} 
(\rho_{1Z} \phi_1^0 - \rho_{2Z} \phi_2^0), \label{eq5}
\eeq 
\beq
\rho_W G_W^{\pm} = \xi_W m_W G_W^{\pm} \equiv 
-(\rho_{1W} H_1^{\pm} - \rho_{2W} H_2^{\pm}), \label{eq6}
\eeq
with parameters $\rho_{iV}$. 
The usual form of the $R_{\xi}$ gauge fixing in the MSSM is 
recovered by the substitution \cite{cpr,gh} 
\beq
(\rho_{1V}, \rho_{2V})=\xi_Vg_V(v_1, v_2)/2=
\xi_Vm_V(\cos\beta, \sin\beta). \label{eq7}
\eeq

The UV divergent corrections to $m_b$ contain one source for  
the SU(2)$\times$U(1) gauge symmetry breaking. It is either 
$v_1$ originated from the shift (\ref{eq2}) of $H_1^0$, 
or $\rho_{1V}$ in the $R_{\xi}$ 
gauge fixing term (\ref{eq4}) and the Fadeev-Popov ghost term. 
The former contribution is obtained from that to the 
$\bar{b}_Rb_L\phi_1^0$ Yukawa coupling $h_b$ by replacing 
external $\phi_1^0$ by $v_1/\sqrt{2}$, 
except for the wave function correction of $H_1^0$ to $h_b$. 
Similar argument holds for the UV divergent corrections to $m_t$ and 
to the $\bar{t}_Rt_L\phi_2^0$ Yukawa coupling $h_t$. As a result, 
if the $\rho_{iV}$ contributions are absent, 
the runnings of $v_i$ are the same as those of the wave functions 
of $H_i^0$, namely 
\bea
\frac{dv_1}{dt} &=& 
\frac{1}{h_b}\left[ \sqrt{2}\frac{d}{dt}(m_b) -\frac{dh_b}{dt}v_1 \right] 
= - \gamma_1 v_1, \nonumber \\
\frac{dv_2}{dt} &=& 
\frac{1}{h_t}\left[ \sqrt{2}\frac{d}{dt}(m_t) -\frac{dh_t}{dt}v_2 \right] 
= - \gamma_2 v_2, \label{eq8}
\eea
where $t\equiv \ln Q_{\dr}$ is the \dr renormalization scale. 
The anomalous dimensions of $H_i^0$ 
are denoted as $\gamma_i$, which generally depend on 
the gauge fixing parameters $\xi$. 
The RGEs (\ref{eq8}) for $v_i$ have been widely used 
in the Landau gauge $\xi=\rho_{iV}=0$. 

However, in general $R_{\xi}$ gauges, $\rho_{iV}$ in the gauge fixing 
terms (\ref{eq4}) may give additional contributions to 
the quark mass running, as 
$\bar{b}b\rho_{1V}$ and $\bar{t}t\rho_{2V}$. Since they have 
no corresponding contributions to the 
$\bar{b}b\phi_1$ and $\bar{t}t\phi_2$ 
couplings, the RGEs for $v_i$ deviate \cite{schilling,cpr} from 
Eq.~(\ref{eq8}). Their general forms are then 
\beq
\frac{dv_i}{dt}= -\gamma_i v_i + Y_{iV} \rho_{iV}, \label{eq9}
\eeq
where $Y_{iV}$ are polynomials of dimensionless couplings. 
Therefore, the RGE for $\tan\beta$ becomes, using Eq.~(\ref{eq7}), 
\beq
\frac{d}{dt}\tan\beta =\tan\beta \left( -\gamma_2 +\gamma_1 
+ \frac{\xi_V g_V}{2}Y_{2V} - \frac{\xi_V g_V}{2}Y_{1V}
\right) . \label{eq10}
\eeq 

I then give explicit form of the RGE for $\tan\beta$ in the MSSM, to the 
two-loop order. First, one-loop RGEs for $v_i$ ($i=1,2$) are 
\bea
\left. 
\frac{dv_i}{dt}\right|_{\rm 1loop} &=& 
-\gamma_i^{(1)}v_i +\frac{1}{(4\pi)^2}(g_Z \rho_{iZ} +2g_2 \rho_{iW}) 
\nonumber \\
&=& v_i \left[ -\gamma_i^{(1)} +\frac{1}{(4\pi)^2}\left(
\frac{\xi_Z g_Z^2}{2} + \xi_W g_2^2 \right) \right]\, , \label{eq11}
\eea
with the one-loop anomalous dimensions $\gamma_i^{(1)}$, 
\beq
(4\pi)^2\gamma_i^{(1)} = N_c h_q^2 
-\frac{3}{4}g_2^2\left( 1-\frac{2}{3}\xi_W-\frac{1}{3}\xi_Z \right) 
-\frac{1}{4}g_Y^2(1-\xi_Z), \label{eq12}
\eeq
where $h_q^2=(h_b^2,h_t^2)$ for $i=(1,2)$, respectively, and $N_c=3$. 
The $\rho_{1Z}$ contribution to $m_b$ is obtained from 
the diagram in Fig.~1. All other contributions of $\rho_{iV}$ to $m_q$ 
come from similar diagrams. Eq.~(\ref{eq11}) is consistent 
with the result in Refs.~\cite{cpr,dabel} for $\xi=1$. 
Since the gauge dependence of $\gamma_i$, as well as the 
contribution from ($\rho_{iZ}$, $\rho_{iW}$) satisfying Eq.~(\ref{eq7}), 
cancels in the ratio (\ref{eq10}), the one-loop running $\tan\beta$ is 
gauge parameter independent in the $R_{\xi}$ gauge. 

I next proceed to the two-loop corrections. 
The two-loop anomalous dimensions $\gamma_i^{(2)}$ are 
obtained from the general formula \cite{2loopNS} in the \ms scheme 
(the modified minimal subtraction schemes with dimensional regularization), 
after conversion into the \dr scheme \cite{mstodr}, as 
\bea
(4\pi)^4\gamma_1^{(2)} &=& -N_c(3h_b^4 + h_b^2h_t^2) 
 +2N_c h_b^2\left( \frac{8}{3}g_3^2 - \frac{1}{9} g_Y^2 \right) +L(g), 
\nonumber \\
(4\pi)^4\gamma_2^{(2)} &=& -N_c(3h_t^4 + h_b^2h_t^2) 
 +2N_c h_t^2\left( \frac{8}{3}g_3^2 + \frac{2}{9} g_Y^2 \right) +L(g). 
\label{eq13}
\eea
The last term $L(g)$ is a gauge-dependent ${\cal O}(g^4)$ polynomial 
and is common both for $\gamma_1^{(2)}$ and $\gamma_2^{(2)}$. 
The $\xi$ dependence of the ${\cal O}(h_q^2g^2)$ terms completely 
cancels out \cite{2loopNS}. 
Note also that the ${\cal O}(h_q^4)$ and ${\cal O}(h_q^2g^2)$ 
terms agree with the result in the $\xi=0$ gauge \cite{RGEvsusy3} 
and with the superfield calculation \cite{2loopS} which uses 
manifestly supersymmetric gauge fixing. 

The two-loop $\rho_{iV}$ contributions to $dv_i/dt$ have 
${\cal O}(h_q^2g\rho_{iV})$ and ${\cal O}(g^3\rho_{iV})$ terms. 
The latter is common for 
both $i=1$ and 2, and cancels out in the ratio $\tan\beta$ if 
Eq.~(\ref{eq7}) is satisfied. Therefore, only the former 
${\cal O}(h_q^2g\rho_{iV})$ contributions are explicitly calculated. 
For example, the ${\cal O}(h_b^2g_Z\rho_{1Z})$ contribution to $v_1$ 
comes from the diagram (a) in Fig.~2, while other diagrams (b,c) 
cancel each other. The RGEs for $v_i$ are finally 
\beq
\left. 
\frac{dv_i}{dt}\right|_{\rm 2loop} = 
-\gamma_i^{(2)} v_i 
- \frac{N_ch_q^2}{(4\pi)^4} (g_Z\rho_{iZ} + 2 g_W \rho_{iW})+P_V(g)\rho_{iV}, 
\label{eq14}
\eeq
where again $h_q^2=(h_b^2,h_t^2)$ for $i=(1,2)$, respectively. 
$P_V(g)$ are possibly gauge-dependent ${\cal O}(g^3)$ functions 
which are common for both $\rho_{1V}$ and $\rho_{2V}$. 
It is therefore seen that, due to the $\rho_{iV}$ contributions 
in Eq.~(\ref{eq14}), the running $\tan\beta$ has 
the ${\cal O}(h_q^2g_2^2,h_q^2g_Y^2)$ gauge parameter dependence. 
Although existing higher-order calculations of the corrections 
to the MSSM Higgs sector \cite{higgs2lEP,higgs2lRG,higgs2lFD,higgs2lCO} 
have not included the contributions of these orders yet, the gauge 
dependence of $\tan\beta$ may cause theoretical problem in future studies 
of the higher-order corrections in the MSSM. 

One way to restore the gauge independence of 
renormalized running $\tan\beta$ is to introduce gauge-dependent 
shifts of $\phi_i^0$ such as to cancel the $\rho_{iV}$ contributions 
to the effective action. 
This modification corresponds to the addition of extra shifts of $v_i$ 
to all diagrams. 
The running $v_i$ in this new definition then obey the same RGEs as 
those for $H_i$, namely Eq.~(\ref{eq8}). 
The modified renormalized 
$\tan\beta$ becomes gauge independent to the two-loop order. 
However, an extra two-loop shift $\delta(v_2/v_1)$ has to be added to 
any quantities which depend on $\tan\beta$. 

Before leaving, I briefly comment on two related issues 
in the process-independent on-shell renormalization of $v_i$ and 
$\tan\beta$ which is used in Refs.~\cite{cpr,dabel}. 
First, they cancel the one-loop $\rho_{iV}$ contributions 
by extra counterterms for $v_i$, $\delta v_i$, and determine 
their finite parts by imposing 
the condition $\delta v_1/v_1=\delta v_2/v_2$. It is clear from 
Eq.~(\ref{eq14}) that this condition has to be modified beyond 
the one-loop. Second, the gauge dependence already 
appears in the one-loop finite part of the 
on-shell counterterm $\delta(\tan\beta)$. This is similar to the 
gauge dependence of the on-shell renormalized mixing matrices 
for other particles \cite{Ugaugedep}. 

In conclusion, I discussed the UV renormalization 
of the ratio $\tan\beta=v_2/v_1$ of the Higgs VEVs in the MSSM, 
to the two-loop order and in general $R_{\xi}$ gauges. 
When renormalized $v_i$ are given by the minimum of the 
loop-corrected effective potential, the contributions of $\rho_{iV}$ in 
the $R_{\xi}$ gauge fixing term cause two-loop gauge dependence 
of the RGE for $\tan\beta$. To avoid this gauge dependence, 
the contributions of $\rho_{iV}$ have to be cancelled 
by extra shift of the Higgs boson fields $\phi_i^0$. 

\vspace{5mm}

{\it Acknowledgements:} 
This work was supported in part by the Grant-in-aid for 
Scientific Research from Japan Society for the Promotion of
Science, No.~12740131. 


\vspace{30mm}

\begin{figure}[htbp]
\begin{center}
\includegraphics[width=17cm]{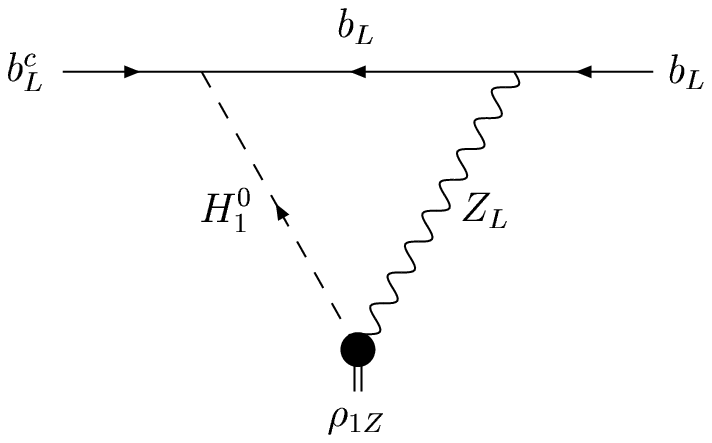}
\end{center}
\caption{ 
One-loop divergent contribution of $\rho_{1Z}$ to $m_b$. There is another 
diagram obtained from this one 
by the interchange $(b_L\leftrightarrow b^c_L)$. 
}
\label{fig1}
\end{figure}

\begin{figure}[htbp]
\begin{center}
\includegraphics[width=17cm]{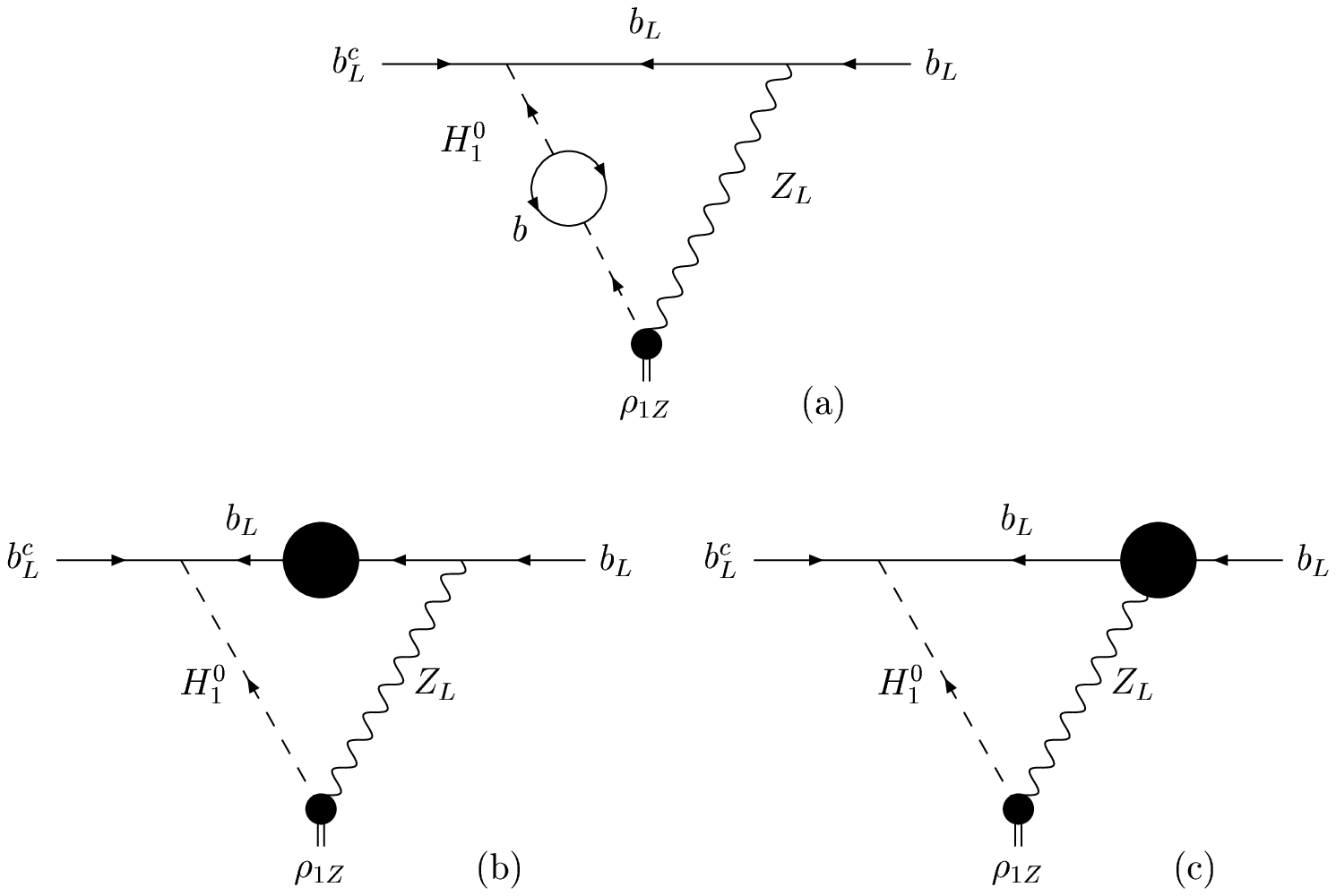}
\end{center}
\caption{ 
Two-loop divergent ${\cal O}(h_q^2g\rho_{1Z})$ contributions to $m_b$. 
The diagrams obtained from (a--c) by the 
interchange $(b_L\leftrightarrow b^c_L)$ also contribute. 
The blobs in diagrams (b, c) denote the ${\cal O}(h_q^2)$ 
quark-Higgs and squark-higgsino subloops. 
One-loop subdivergences are subtracted in the calculation. 
The divergences of (b) and (c) completely cancel out. 
}
\label{fig2}
\end{figure}


\baselineskip=14pt

\begin{thebibliography}{99}

\bibitem{mssm}
H. P. Nilles, Phys. Rep. {\bf 110} (1984) 1;\\
H. E. Haber and G. L. Kane, Phys. Rep. {\bf 117} (1985) 75;\\
R. Barbieri, Riv. Nuov. Cim. {\bf 11} (1988) 1;\\
S. P. Martin, hep-ph/9709356, in {\it Perspectives on Supersymmetry}, 
edited by G.L. Kane (World Scientific, 1998). 

\bibitem{gh} 
J. F. Gunion and H. E. Haber, Nucl. Phys. {\bf B272} (1986) 1; 
{\bf B402} (1993) 567(E). 

\bibitem{sirlin}
A. Sirlin, Phys. Rev. D {\bf 22} (1980) 971. 

\bibitem{onshell}
K.-I. Aoki, Z. Hioki, R. Kawabe, M. Konuma, and T. Muta, 
Prog. Theor. Phys. Suppl. {\bf 73} (1982) 1;\\
M. B\"ohm, H. Spiesberger, and W. Hollik,
Fortschr. Phys. {\bf 34} (1986) 687;\\
A.~Denner, Fortschr. Phys. {\bf 41} (1993) 307.

\bibitem{tanbrun}
A. Yamada, Phys. Lett. B {\bf 263} (1991) 233; 
Z. Phys. C {\bf 61} (1994) 247; \\
A. Brignole, Phys. Lett. B {\bf 281} (1992) 284;\\
D. Pierce and A. Papadopoulos, Phys. Rev. D {\bf 47} (1993) 222;\\
H. E. Haber and R. Hempfling, Phys. Rev. D {\bf 48} (1993) 4280.

\bibitem{cpr}
P. H. Chankowski, S. Pokorski, and J. Rosiek, 
Phys. Lett. B {\bf 274} (1992) 191; 
Nucl. Phys. {\bf B423} (1994) 437; 497. 

\bibitem{dabel}
A. Dabelstein, Z. Phys. C {\bf 67} (1995) 495; 
Nucl. Phys. {\bf B456} (1995) 25.

\bibitem{dpf}
Y. Yamada, hep-ph/9608382, in {\it DPF '96: The Minneapolis Meeting}, 
edited by H. Keller, J. K. Nelson, and D. Reeder (World Scientific, 1998). 

\bibitem{DR}
W. Siegel, Phys. Lett. {\bf 84B} (1979) 193;\\
D. M. Capper, D. R. T. Jones, and P. van Nieuwenhuizen, 
Nucl. Phys. {\bf B167} (1980) 479.

\bibitem{rxi}
K. Fujikawa, B. W. Lee and A. I. Sanda, Phys. Rev. D {\bf 6} (1972) 2923. 

\bibitem{RGExi}
D. J. Gross and F. Wilczek, Phys. Rev. D {\bf 8} (1973) 3633;\\
W. E. Caswell and F. Wilczek, Phys. Lett. {\bf 49B} (1974) 291;\\
H. Kluberg-Stern and J. B. Zuber, 
Phys. Rev. D {\bf 12} (1975) 467; 482.

\bibitem{degrassi}
G. Degrassi and A. Sirlin, Nucl. Phys. {\bf B383} (1992) 73;\\
R. Hempfling and B. A. Kniehl, Phys. Rev. D {\bf 51} (1995) 1386.

\bibitem{nielsenSM}
P. Gambino and P. A. Grassi, Phys. Rev. D {\bf 62} (2000) 076002. 

\bibitem{mnuR}
M. Hirsch, M. A. D\'{\i}az, W. Porod, J. C. Rom\~ao, and J. W. F. Valle, 
Phys. Rev. D {\bf 62} (2000) 113008. 

\bibitem{neumass}
P. H. Chankowski and P. Wasowicz, hep-ph/0110237. 

\bibitem{RGEvsusy}
G. Gamberini, G. Ridolfi, and F. Zwirner, Nucl. Phys. {\bf B331} (1990) 331. 

\bibitem{veff1}
R. Jackiw, Phys. Rev. D {\bf 9} (1974) 1686;\\
L. Dolan and R. Jackiw, Phys. Rev. D {\bf 9} (1974) 2904.

\bibitem{veff2}
N. K. Nielsen, Nucl. Phys. {\bf B101} (1975) 173.

\bibitem{veff3}
I. J. R. Aitchison and C. M. Fraser, Ann. Phys. (N.Y.) {\bf 156} (1984) 1;\\
D. Johnston, Nucl. Phys. {\bf B253} (1985) 687; {\bf B283} (1987) 317. 

\bibitem{cima}
O. M. Del Cima, D. H. T. Franco, and O. Piguet, 
Nucl. Phys. {\bf B551} (1999) 813.

\bibitem{schilling}
M. Okawa, Prog. Theor. Phys. {\bf 60} (1978) 1175;\\
A. Schilling and P. van Nieuwenhuizen, Phys. Rev. D {\bf 50} (1994) 967. 

\bibitem{2loopNS}
M. E. Machacek and M. T. Vaughn, Nucl. Phys. {\bf B222} (1983) 83. 

\bibitem{mstodr}
S. P. Martin and M. T. Vaughn, Phys. Lett. B {\bf 318} (1993) 331.

\bibitem{RGEvsusy3}
S. P. Martin, hep-ph/0111209. 

\bibitem{2loopS}
P. West, Phys. Lett. B {\bf 137} (1984) 371;\\
D. R. T. Jones and L. Mezincescu, Phys. Lett. B {\bf 138} (1984) 293;\\
Y. Yamada, Phys. Rev. D {\bf 50} (1994) 3537. 

\bibitem{higgs2lEP}
R. Hempfling and A. H. Hoang, Phys. Lett. B {\bf 331} (1994) 99;\\
R. Zhang, Phys. Lett. B {\bf 447} (1999) 89. 

\bibitem{higgs2lRG}
M. Carena, M. Quir\'os, and C. E. M. Wagner, 
Nucl. Phys. {\bf B461} (1996) 407;\\
H. E. Haber, R. Hempfling, and A. H. Hoang, Z. Phys. C {\bf 75} (1997) 539. 

\bibitem{higgs2lFD}
S. Heinemeyer, W. Hollik, and G. Weiglein, 
Phys. Rev. D {\bf 58} (1998) 091701; Phys. Lett. B {\bf 440} (1998) 296; 
Eur. Phys. J. C {\bf 9} (1999) 343; {\bf 16} (2000) 139. 

\bibitem{higgs2lCO}
J. Espinosa and R. Zhang, JHEP {\bf 0003} (2000) 026; 
Nucl. Phys. {\bf B586} (2000) 3;\\
M. Carena, H. E. Haber, S. Heinemeyer, W. Hollik, C. E. M. Wagner, 
and G. Weiglein, Nucl. Phys. {\bf B580} (2000) 29;\\
G. Degrassi, P. Slavich, and F. Zwirner, Nucl. Phys. {\bf B611} (2001) 403;\\
A. Brignole, G. Degrassi, P. Slavich, and F. Zwirner, hep-ph/0112177. 

\bibitem{Ugaugedep}
P. Gambino, P. A. Grassi, and F. Madricardo, 
Phys. Lett. B {\bf 454} (1999) 98;\\
B. A. Kniehl, F. Madricardo, and M. Steinhauser, 
Phys. Rev. D {\bf 62} (2000) 073010;\\
A. Barroso, L. Br\"ucher, and R. Santos, 
Phys. Rev. D {\bf 62} (2000) 096003;\\ 
Y. Yamada, Phys. Rev. D {\bf 64} (2001) 036008. 

\end{thebibliography}
\end{document}